\documentclass[graybox]{svmult}

\usepackage{mathptmx}       
\usepackage{helvet}         
\usepackage{courier}        
\usepackage{type1cm}        
\usepackage{makeidx}         
\usepackage{graphicx}        
\usepackage{multicol}        
\usepackage[bottom]{footmisc}
\usepackage{multirow}
\usepackage{cite}

\makeindex             

\begin{document}

\title*{DSARSR: Deep Stacked Auto-encoders Enhanced Robust Speaker Recognition}
\author{Zhifeng Wang, Chunyan Zeng, Surong Duan, Hongjie Ouyang, and Hongmin Xu}
\institute{Zhifeng Wang, Surong Duan, Hongjie Ouyang, and Hongmin Xu \at Department of Digital Media Technology, Central China Normal University, Wuhan, Hubei, China, 430079
\and Correspondence author: Chunyan Zeng \at Hubei Key Laboratory for High-efficiency Utilization of Solar Energy and Operation Control of Energy Storage System, Hubei University of Technology, Wuhan, Hubei, China, 430068,
 \email{cyzeng@hbut.edu.cn}}
%
%
\maketitle

\abstract*{}

\abstract{Speaker recognition is a biometric modality which utilize speaker’s speech segments to recognize identity, determining whether the test speaker belongs to one of the enrolled speakers. In order to improve the robustness of i-vector framework on cross-channel conditions and explore the nova method for applying deep learning to speaker recognition, the Stacked Auto-encoders is applied to get the abstract extraction of the i-vector instead of applying PLDA. After pre-processing and feature extraction, the speaker and channel independent speeches are employed for UBM training. The UBM is then used to extract the i-vector of the enrollment and test speech. Unlike the traditional i-vector framework, which uses linear discriminant analysis (LDA) to reduce dimension and increase the discrimination between speaker subspaces, this research use stacked auto-encoders to reconstruct the i-vector with lower dimension and different classifiers can be chosen to achieve final classification. The experimental results show that the proposed method achieves better performance than the-state-of-the-art method.}

\section{Introduction}

Speaker recognition is a biometric modality which utilize speaker’s speech segments to recognize identity, determining whether the test speaker belongs to one of the enrolled speakers \cite{Wang2021m,Zeng2018,Wang2020h}. Speaker recognition can be regarded as an means of identification which is of great use for application like forensics, transaction authentication as well as law enforcement \cite{Wang2011a,Wang2011,Zhu2013}. The development of speaker recognition technology has gone through the following four stages.

The first stage was from the 1960s to the 1970s, when the research was focused on speech feature extraction and template matching techniques. In 1962, Bell Labs proposed a method which use the spectrogram to recognize speakers \cite{Atal1971Speech}. After that Atal et al. proposed Linear Predictive Cepstrum Coefficient (LPCC) \cite{patent:3700815}, which improved the accuracy of speaker recognition.

The second stage was from the 1980s to the mid-1990s, when the statistical model began to be applied into speaker recognition. Mel-frequency cepstrum (MFCC) was represented by Davis \cite{Ieee1990Comparison}, which is a representation of the short-term power spectrum of a speech signal.

Around 2000, the GMM-UBM framework for text-independent speaker recognition proposed by Reynolds reduced the GMM's dependence on the training set \cite{Reynolds2000Speaker}. In 2006, Campbell proposed the Gaussian mixture super vector-support vector machine model (GSV-SVM) \cite{Campbell02generalizedlinear} based on GMM-UBM and support vector machine becoming the predominant technologies. After 2010, models like joint factor analysis (JFA) \cite{Kenny05jointfactor}, and i-vector\cite{dehak2011front} based on Gaussian super vector made the enormous promotion for speaker recognition system. Based on i-vector, Kenny was inspired by the conventional linear discriminant analysis (LDA) \cite{Mclaren2011Source} for face recognition and proposed Probabilistic Linear Discriminant Analysis (PLDA) \cite{Prince2007Probabilistic}, which is the probabilistic form of LDA. 

The fourth stage began in this century (2010) when deep learning began to be introduced into speaker recognition. For speaker verification task, deep models are employed both in feature extraction (such as Deep RBMs, Speaker-discriminant DNN) and training phase. For speaker identification task, bottleneck (BN) features were proposed for nonlinear feature transformation and dimensionality reduction \cite{Song2015Deep} \cite{Matejka14neuralnetwork}. \cite{Zhang2015} presented DAE-based dereverberation for feature extraction and built a robust distant-talking speaker recognition method.

This research is inspired by the great success of deep learning in computer vision \cite{Zeng2023c,Li2023,Zeng2022,Wang2022at,Zeng2021c,Wang2021,Zeng2020a,Min2018,Wang2017}, data mining \cite{Wang2023,Lyu2022}, speech processing \cite{Zeng2023b,Zeng2023,Wang2023a,Zeng2022a,Wang2022t,Zeng2021a,Zeng2021b,Zeng2020,Wang2018a,Wang2015b}, and other areas \cite{Wang2023d,Li2023a,Wang2022as,Min2019,Tian2018a,Wang2015a}. In order to improve the robustness of i-vector framework on cross-channel conditions and explore the possible direction for applying deep learning to speaker recognition, instead of applying PLDA we intend to use the Stacked Auto-encoders to get the abstract extraction of the i-vector and classifiers like SVM and neural network to do the final classification, which is new to this field.

\section{The Framework for Robust Speaker Recognition based on Stacked Auto-encoders}
The basic framework of the Stacked Auto-encoders based speaker recognition model can be illustrated in Fig. 1. After pre-processing and feature extraction, the speaker and channel independent speech are employed for UBM training. The UBM is then used to extract the i-vector of the enrollment and test speech. Unlike the traditional i-vector framework, which uses linear discriminant analysis (LDA) to reduce dimension and increase the discrimination between speaker subspaces, we use stacked auto-encoders to reconstruct the i-vector with lower dimension and different classifiers can be chosen to achieve final classification.

\begin{figure}[h]
	\centering
	\includegraphics[width=5in,height=2in]{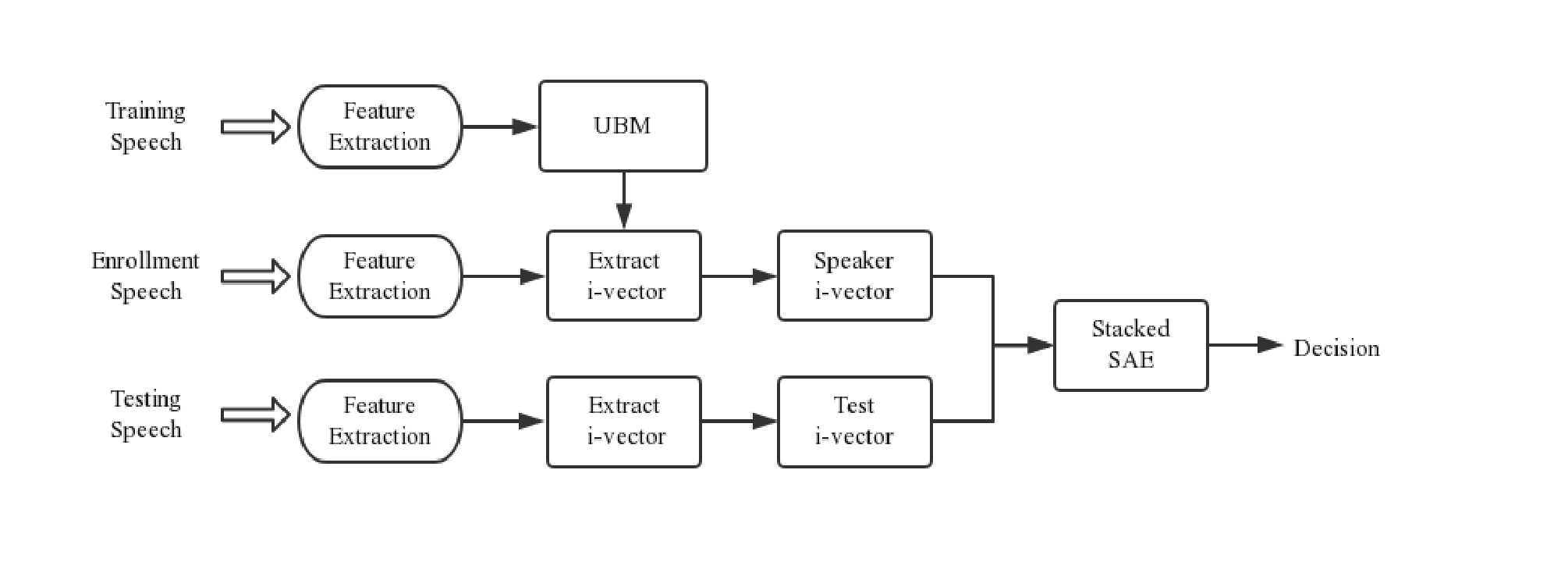}
	\caption{Stacked Auto-encoders based Speaker Recognition Model }
\end{figure}

\subsection{I-vector Extraction for Speaker Recognition}
The i-vector framework is now considered as the state-of-the-art for speaker recognition, firstly proposed in 2009 \cite{dehak2011front}. This algorithm combines the strengths of GMM supervector SVMs \cite{1618704} and Joint Factor Analysis (JFA) \cite{Kenny05jointfactor} which introduces the total variability space (T) containing the differences not only between the speakers but also between the channels.

The i-vector framework can be illustrated as below. After feature extraction, the Universal Background Model (UBM) is obtained using EM. Then the zero and first order Baum Welch Statistics can be computed as:
$$N_{c}^{k}=\sum_{t}\gamma_{c,t}^{k}$$
$$F_{c}^{k}=\sum_{t}\gamma_{c,t}^{k}o_{t}^{k}$$

Given the statistics above, the T matrix is trained using the maximum likelihood estimate (MLE). 

Step E: Randomly initialize the total variability matrix T before training. Then calculate the variance and mean of the speaker factor
$$l_{T}(S)=I+T^{T}\Sigma^{-1}NN(s)T$$
$$y(S)=l_{T}^{-1}(s)T^{T}\Sigma^{-1}FF(s)$$

Step M: Maximum likelihood revaluation. The statistics of all the enrollment data was added:
$$N_{c}=\sum_{s}N_{c}(s)$$
$$A_{c}=\sum_{s}N_{c}(s)l_{T}^{-1}(s)$$
$$C=\sum_{s}FF(s)y(s)^{T}$$

After obtaining matrix T, i-vector can be extracted. The processes of extraction are: First, calculating the Baum-Welch statistic of the corresponding speaker then the estimation of i-vector using matrix T can be calculated as:
$$E[w_{s,k}]=(I+T^{T}\Sigma^{-1}N_{h}(s)T )^{-1}T^{T}\Sigma^{-1}F_{h}(s)$$

Where $\Sigma$ is the covariance matrix of the UBM. In general, the dimension of i-vector range from 400 to 600.

After obtaining the initial i-vector, linear discriminate analysis (LDA) with Fisher criterion is normally used to reduce the dimensionality (typical dimensions are 200 dimensions) as well as increase the discrimination between speaker subspaces. Then, the within class covariance normalization (WCCN) is performed so that the speaker subspace can be orthogonal. Finally, to score the verification trails, the log-likelihood ratio (LLR) was computed between same (H1) versus different speakers hypotheses (H0):
$$llr= \ln \frac{p(x_{1},x_{2}|H_{1})}{p(x_{1}|H_{0}) p(x_{2}|H_{0})}$$

\subsection{Stack Auto-encoders for Robust Speaker Recognition}
As early as 1986, Rumelhart proposed the auto-encoder \cite{article2} for dimensionality reduction of high-dimensional data. After the improvement by Hinton, the concept of deep auto-encoder is proposed which is an unsupervised learning for nonlinear feature extraction mainly used for data representation and reconstruction. The  most outstanding feature of auto-encoders is its link to latent variable space, 
which make it special among generative model.

The main objective of the auto-encodes is to learn the internal representation of the input data with a more compact form, which means to extract the essences of the data while losing the minimal useful information. The model aims to give priority to the significant parts of the original data to learn about data. Architecturally, the basic framework of an auto-encoders is a feed-forward network similar to MLP but with different purpose, illustrated in Fig. 2. Instead of getting a prediction output of the input data, auto-encoders aims in reconstructing the input data to get a low dimensional representation.

\begin{figure}[h]
	\centering
	\includegraphics[width=3in,height=2in]{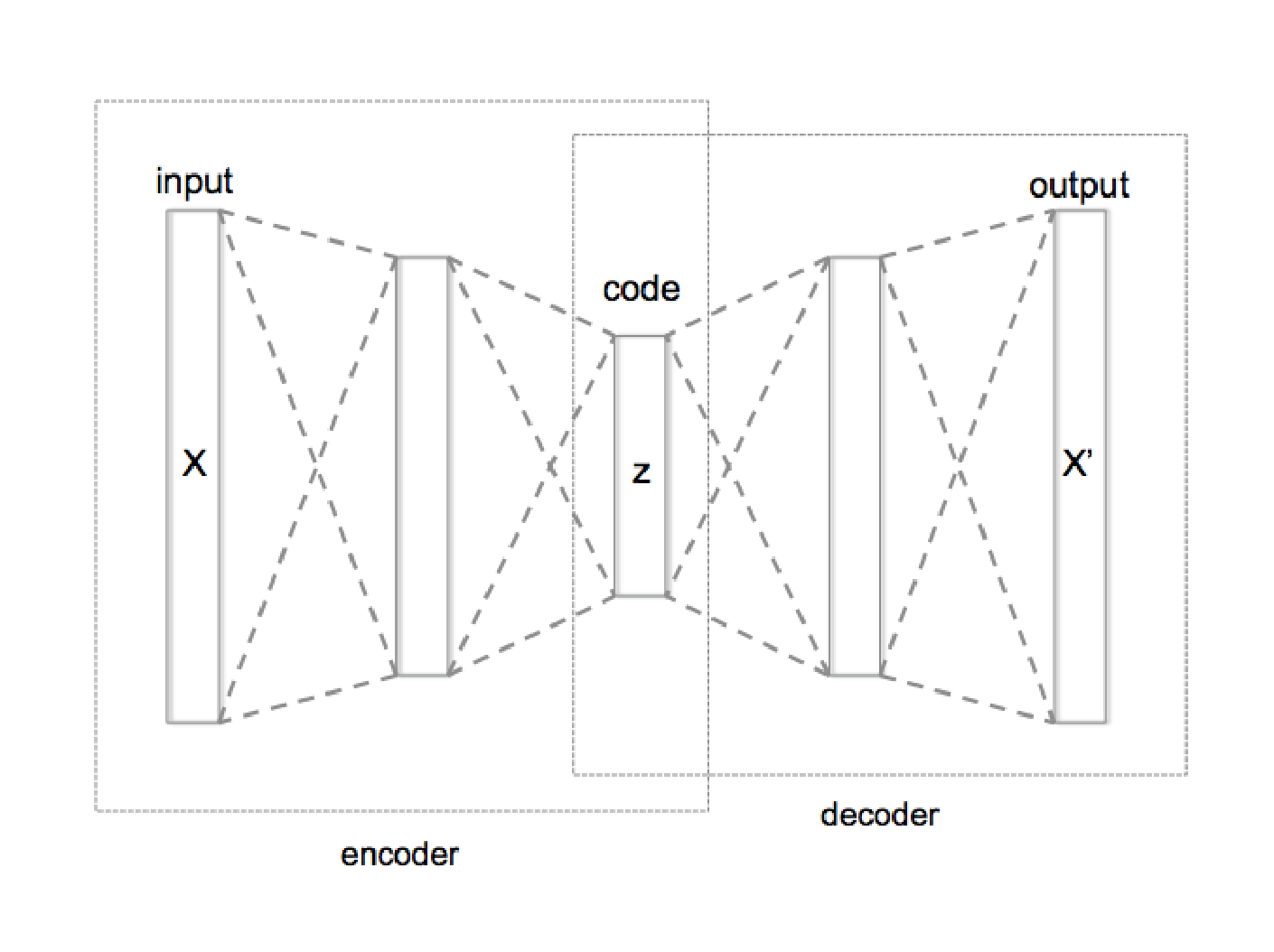}
	\caption{The Designed Structure of Auto-encoder for Speaker Recognition}
\end{figure}

The auto-encoder always includes two parts -- encoder $\phi$  and the decoder $\psi$ which can be defined as transitions between feature space with different dimension:

$$\phi:\textit{X}\rightarrow \textit{F}$$
$$\psi:\textit{F} \rightarrow \textit{X}$$
$$\phi , \psi=\arg \min_{\phi , \psi}||X-(\psi\circ \phi)X||^{2}$$

During the encoder phase, the input $x$ is mapped to code $z\in R^{P}=F$:
$$z=\sigma(Wx+b)$$

As for decoder phase, $z$ is mapped backwards to input space to reconstruct $x$:
$$x^{'}=\sigma^{'}(W^{'}z+b^{'})$$

The object function of auto-encoders is to minimize the reconstruction errors in order to restore input $x$:
$$L(x,x^{'})=||x-x^{'}||^{2}=||x-\sigma^{'}(W^{'}(\sigma(Wx+b))+b^{'})||^{2}$$

A stacked auto-encoder network contains multiple hidden layers and is symmetric about the intermediate layer, containing one input layer and corresponding output layer with $2r-1$ hidden layers. Suppose $m$ nodes of input layerdenoted as $ x=(x_{1},x_{2},...,x_{m})^{T}\in R^{m}$, the hidden layer vector is $h_{k}=(h_{k,1},h_{k,2},...,h_{k,n_{k}})^{T} \in R^{n_{k}}$, and the output vector is $ {x}'=({x}'_{1},{x}'_{2},...,{x}'_{m})^{T}\in R^{m}$ then each hidden layers of the auto-encoders can be denoted as:
$$h_{1}=\sigma_{h_{1}}(W^{1}x+b^{1})$$
$$h_{k}=\sigma_{h_{k}}(W^{k}h_{k-1}+b^{k}),2 \leq k \leq 2r-1$$
$${x}'={\sigma}'_{h_{k}}(W^{2r}h_{2r-1}+b^{2r})$$

The hidden layer code can be regarded as the compression of the input space if it has lower dimensionality. By connecting multiple similar encoders, the output of the$ n_{th}$ layer is regarded as the input of the $n + 1$ layer. After multi-layer training, the auto-encoder can extract the essence features from the original data, and then construct another neural network of or add a classifier such as SVM or LR, the classification can be efficiently implemented.

We want to have the auto-encoder with the capacity of learning the useful properties of the input data and it can be achieved by constraining the hidden layer to have a smaller dimension than the input layer. An auto-encoder with lower dimension of the hidden layer is called undercomplete. The undercomplete architecture makes the auto-encoders to capture the most significant features of the input data.

If the decoder $\varphi $  is linear and the loss function is the mean squared error, the auto-encoder is just like PCA which is to learn the principal subspace of the input data. 
Auto-encoders with nonlinear encoder and decoder has much stronger generalization capacity than PCA. However, the strong learning capacity of the encoder and decoder can easily 
lead to over-fitting without extracting useful knowledge from input.

\section{Experiments and Results}
\subsection{Database Description}
Two databases are applied in this thesis: the TIMIT corpus and the 2006 NIST Speaker Recognition Evaluation Training Set. We employed 354 speakers from TIMIT, with 3540 speech utterances in total, from which 2000 were used for training the UBM, the remaining 1540 for i-vector extraction. And for the NIST 2006 database, we employed 400 speakers in total and 300 for modeling the UBM and 100 for constructing the speaker recognition system. Each speaker has eight two-channel (4-wire) conversations collected over telephone channels which are mostly in English and four other languages. In this work, our experiments can be divided into two main parts based on TIMIT and NIST 2006 databases, in terms of the gender detection and speaker recognition.

\subsection{Speaker Gender Detection}
As to prove the feasibility of this method, we first apply it to achieve a binary classification task -- speaker gender detection. Speaker gender detection is not necessarily considered as an end itself, however it can be used as a pre-processing step in Automatic Speech Recognition, allowing the selection of gender dependent acoustic models \cite{Lamel1995A} \cite{Bocklet2008Age}. In recent research, this system has been used for selecting gender specific emotion recognition engines \cite{Xia2014Modeling} and defining different strategies for different gender \cite{Shafey2014Audio}.

\subsubsection{Evaluation of Performance}
In this experiment, ACC, AUC and MCC are employed for evaluate the performance of the model which are the common evaluation method for binary classification.

The ROC curve is used as a measurement for classifier model based on TPR and FPR.
The AUC (Area Under Curve) is an evaluation index often used in the binary classification model, defined as the area under the ROC curve. The TPR and FPR can be denoted as below:
$$ FPR=\frac{FP}{FP+TN}$$
$$TPR=\frac{TP}{TP+FN}$$
Then the classification accuracy can be defined as:
$$ACC=\frac{TP+TN}{FP+TN+TP+FN}$$

Where TP -- True Positives, FP -- False Positives, FN -- False Negatives, TN -- True Negatives. The ROC curve can keep itself unchanged when the distribution of positive and negative samples changes. When classification is completely random, the AUC is close to 0.5, and the closer the value of AUC is to 1, the better the model prediction effect is.

The Matthews correlation coefficient is another measurement usually used to evaluate the binary classifier model, calculated as:
$$MCC=\frac{TP\times TN -FP \times FN}{\sqrt{(TP+FP)(TP+FN)(TN+FP)(TN+FN)}}$$

The range  is $[-1,1]$, $MCC=-1$ denoting the worst prediction, $0$ denoting the random prediction and $1$ denoting the best prediction.

\subsubsection{Results and Analysis}
In this experiment, the method presented above is applied to the TIMIT database.  108 males and 92 females are selected for UBM training, each speaker includes 10 speech signals of 10s, and 12-dimensional MFCC are extracted by 256 frames. After normalization, the UBM with 64 Gauss components is trained. To extract the i-vector, 77 males and 77 females (different from the data used for training the UBM) are employed. Then the zero and first order Baum Welch Statistics are computed and a total variability space is learned, which was applied to extract the i-vector with 400 dimension.

To reconstruct the i-vector, we apply one layer (200 nodes) and two layers (200 and 40 nodes respectively) of auto-encoder network respectively. Then we apply the Support Vector Machine (SVM) and two layers neural networks as the back-end classifier. The results are as follows.

\begin{table}
	\caption{Experiment Result}
	\centering
	\label{table:result}
	\begin{tabular}{l c c c c}
		\hline 
		\multirow{2}{*}{Measurement} & \multicolumn{2}{c}{One layer auto-encoder} & \multicolumn{2}{c}{Two layers auto-encoder} \\ 
		\cline{2-5}
		& SVM based & Neural network  & SVM based & Neural network  \\ 
		\hline
		ACC & 84\% &  96\% &  78\%  &  98.4\%  \\
		
		AUC & 0.931 & 0.995 & 0.823  & 0.432 \\
		
		MCC & 0.694 & 0.910 & 0.989 & 0.961 \\
		\hline 
	\end{tabular}
\end{table}

In this case, applying two layers of auto-encoders and neural network as the back-end classifier achieved a better classification performance with $98.4\%$ accuracy. As for SVM classifier, it may lead to overfitting problem since the accuracy on the training set is much better than testing set.

\subsection{Speaker Verification}
We applied the undercomplete auto-encoders to NIST 2006 challenge. The basic flowchart of the method is similar with the one shown in 3.4.2. We employed 400 speakers in total and 300 for modeling the UBM and 100 for constructing the speaker recognition system. VAD and feature scaling are used for pre-processing. One Hot Encoder is used for converting the categorical label into numerical label. For this multiple classification task, we employ the recognition rate and confusion matrix to evaluate our model.

\subsubsection{Results and comparison}
In this experiment, a 64 diagonal component UBM is trained, along with a 400 dimensional i-vector extractor. Firstly, we use undercomplete auto-encoders to reconstruct the feature with two hidden layers, which have 200 and 50 nodes respectively. After the final classification, we achieved $91.25\%$ recognition rate and the confusion matrix of the results is shown below.

\begin{figure}[h]
	\centering
	\includegraphics[width=3in,height=2.5in]{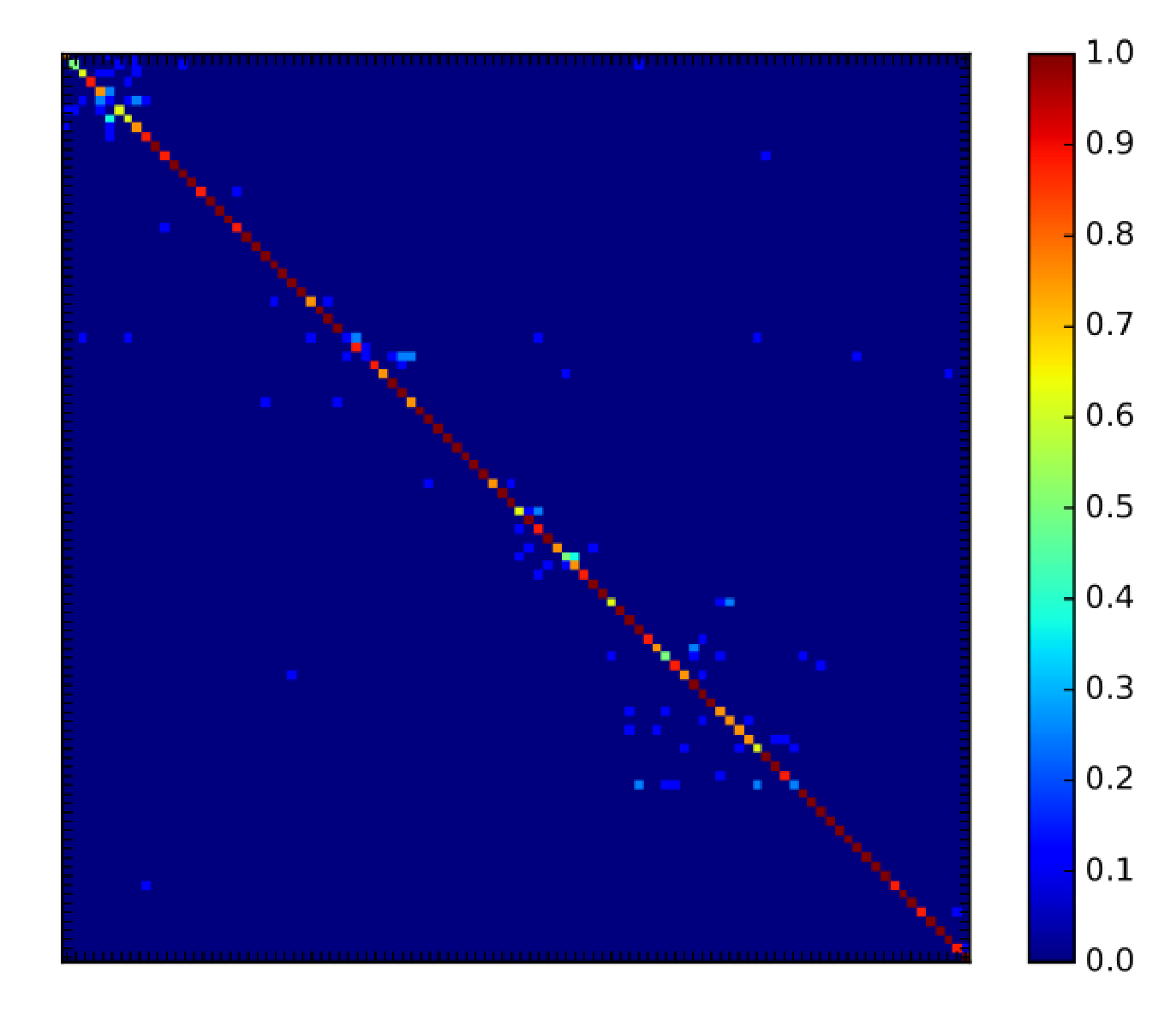}
	\caption{Confusion Matrix based on Undercomplete Auto-encoders Method}
\end{figure}

As the figure shows, the simply used auto-encoders may not achieve a good performance in the final classification task while the reconstruction loss is small enough, which indicate that the undercomplete auto-encoders may fail to learn the useful representation of the input data when given to strong learning ability.

\section{Conclusion}
In this section, we reviewed the general concept of auto-encoders and its basic form -- undercomplete auto-encoders. We intend to take the advantages of the strong capability of represent the feature and replace the LDA in the traditional i-vector frame to realize dimensionality reduction. We employed the proposed model to deal with two classification task – gender detection and speaker recognition and we also explore the performance of different back-end classifiers after using stacked auto-encoders for feature extraction. The performance of undercomplete auto-encoders is not quite good and it seems to lose some useful knowledge of the input data when given to strong learning ability. In next section, we will explore more variant of auto-encoders and compare the performance on the same dataset with the original model.

\begin{acknowledgement}
This research was supported by National Natural Science Foundation of China (No.61901165, 61501199), Science and Technology Research Project of Hubei Education Department (No. Q20191406), Hubei Natural Science Foundation (No. 2017CFB683), and self-determined research funds of CCNU from the colleges’ basic research and operation of MOE (No. CCNU20ZT010).
\end{acknowledgement}

\bibliographystyle{spmpsci}
\bibliography{references,Citations}
\end{document}